\documentclass{jps-cp}
\usepackage{txfonts} %Please comment out this line unless the txfonts package is availabe in your LaTeX system.
\usepackage{graphicx}
\usepackage{xcolor}

\usepackage{amsmath}
\usepackage{url}
\usepackage[symbol]{footmisc}
\title{ Arbitrage impact on the relationship between XRP price and correlation tensor spectra of transaction networks}
\author{Abhijit \textsc{Chakraborty }$^{1,2,\dotplus}$\thanks{Currently at the Department of Humanities and Social Sciences, Indian Institute of Science Education and Research Tirupati, Tirupati 517619, India.} 
and
Yuichi \textsc{Ikeda}$^{1,\dag}$}
\inst{$^{1}$ Kyoto University, Graduate School of Advanced Integrated Studies in Human Survivability, Kyoto, 606-8306, Japan \\
$^{2}$RIKEN Interdisciplinary Theoretical and Mathematical Sciences Program, Saitama, 351-0198, Japan}

\email{ $\dotplus$ abhijit@iisertirupati.ac.in\\
$\dag$ ikeda.yuichi.2w@kyoto-u.ac.jp
}

\abst{
The increasing use of cryptoassets for international remittances has proven to be faster and more cost-effective, particularly for migrants without access to traditional banking. However, the inherent volatility of cryptoasset prices, independent of blockchain-based remittance mechanisms, introduces potential risks during periods of high volatility. This study investigates the intricate dynamics between XRP price fluctuations across diverse crypto exchanges and the correlation of the largest singular values of the correlation tensor of XRP transaction networks.
Particularly, we show the impact of arbitrage opportunities across different crypto exchanges on the relationship between XRP price and correlation tensor spectra of transaction networks.
Distinct periods, non-bubble and bubble, showcase different characteristics in XRP price fluctuations. Establishing a connection between XRP price and transaction networks, we compute correlation tensors and singular values, emphasizing the significance of the largest singular value. Comparisons with reshuffled and Gaussian random correlation tensors validate the uniqueness of the empirical tensor. A set of simulated weekly XRP prices, resembling arbitrage opportunities across various crypto exchanges, further confirms the robustness of our findings. It reveals a pronounced anti-correlation during bubble periods and a non-significant correlation during non-bubble periods with the largest singular value, irrespective of price fluctuations across different crypto exchanges.}
\kword{cryptoassets, price fluctuation, correlation tensor, singular values}

\begin{document}
%\footnote[4]{Currently at the Department of Humanities and Social Sciences, Indian Institute of Science Education and Research Tirupati, Tirupati 517619, India.}
\maketitle

\section{Introduction}
In recent years, cryptoassets have become widely used as an essential means of international remittances~\cite{mills2016distributed, rella2019blockchain}. For example, when migrants transfer money earned from working in the destination country to family members in the country of origin, international remittances using cryptoassets are faster and cheaper than the traditional method of transferring money via banks. In particular, for migrants who do not have bank accounts in their economically developing countries of origin, international remittances using cryptoassets are the only practical way to send money. However, the prices of cryptoassets are more volatile than fiat currency at times, and international remittances during periods of high price volatility are said to be risky~\cite{katsiampa2019empirical}. The price of cryptoassets is independent of the blockchain-based remittance mechanism for cryptoassets. The price is determined when fiat currency and cryptoassets are exchanged on cryptoasset exchanges.

The price of a cryptoasset generally varies from exchange to exchange, as well as the point in time when the cryptoasset is exchanged for fiat currency. Concerning differences in the price of cryptoassets at different points in time, if cryptoassets are purchased and remitted abroad to family members in the country of origin when the price of cryptoassets is low, and if cryptoassets can be sold when the price of cryptoassets is high, the family members in the country of origin can obtain more economic value. On the other hand, concerning the price difference between exchanges, if cryptoassets are purchased on an exchange where the price of cryptoassets is low and remitted abroad to family members in the country of origin, and cryptoassets are sold on an exchange where the price of cryptoassets is high, the family members in the country of origin can obtain more economic value. This is referred as an arbitrage.

The application of Random Matrix Theory (RMT)~\cite{mehta2004random, potters2020first, sengupta1999distributions} has become a prevalent approach in the analysis of financial time series data, encompassing domains like stock prices~\cite{laloux1999noise, plerou1999universal, plerou2002random}, foreign exchange rate~\cite{chakraborty2018deviations, chakraborty2020uncovering}, and other macroeconomic indicators. Recently, we utilized random matrix theory to study crypto transaction networks~\cite{chakraborty2023projecting, chakraborty2023dynamic, chakraborty2023embedding, ikeda2022characterization, ikeda2023hodge}.  Specifically, our method~\cite{chakraborty2023projecting, chakraborty2023dynamic} involving correlation tensor spectra provides a crucial understanding of the relationship between XRP price and the largest singular value.
Our previous work~\cite{chakraborty2023projecting, chakraborty2023dynamic} found that the largest singular value of the correlation tensor for the transaction networks and the price of cryptoassets (on a particular exchange) shows a significant negative correlation. Using this negative correlation may, therefore, reduce the risk of price volatility of cryptoassets at different points in time. However, no attention has so far been paid to the impact of exchanges on the price difference of cryptoassets. If the price difference by exchange is more significant than the difference by point in time, then the negative correlation described above would have no practical significance.

In this paper, we first examine the time series of the price of the cryptoasset XRP on representative exchanges to determine the magnitude of the price differences by the exchange. It then examines the correlation between the first singular value of the correlation tensor of the transaction network and the price, considering the exchange-specific price differences obtained from the actual price time series. Statistical tests test the significance of the correlation. This study aims to answer whether the negative correlations between the largest singular value of the correlation tensor for the transaction networks and the prices of cryptoassets obtained in our previous studies are statistically significant when considering the differences in prices between exchanges.

The paper is organized as follows. We describe our data in Section~\ref{data}.  In Section~\ref{method} we provide a brief description of our methods. Following that, Section~\ref{result} presents a comprehensive discussion of the results obtained from our investigation. Finally, in Section~\ref{conclusion}, we share concluding observations.

\section{Data}
\label{data}
We collect XRP daily close price data from nine different crypto exchanges. The typical 24-hour trade volume on these crypto exchanges is tabulated in Table~\ref{t1}. It is important to note that among the nine exchanges, Bittrex and FTX are currently not operational. We additionally present the monthly volumes across three leading cryptoasset exchanges in Fig.~\ref{f7}.

XRP transaction data is collected from the Ripple API. We construct weekly networks of users' wallets from the transactional data by aggregating the XRP transaction volume for a week between pairs of wallets. It is noteworthy that the weekly transaction networks are weighted and directed in nature. XRP wallets serve as the nodes, and a weighted directed link represents the total XRP transferred from a source wallet to a destination wallet in a week.

\begin{table}[tbh]
\caption{Crypto exchanges in terms of  24 hour volume on February7, 2024, according to analytics website Coinranking.}
\label{t1}
\begin{center}
\begin{tabular}{lr}
Exchange & 24h Volume ( billion USD)\\
\hline
Binance & 11.20 \\
OKX & 1.39 \\
%Investing & DDD \\
Huboi & 1.20 \\
Kraken & 0.85 \\
Bitstamp & 0.19 \\
Bitfinex  & 0.18 \\
Exmo & 0.03 \\
Bittrex & - \\
FTX & - \\
%Coin market cap & DDD \\
\hline
\end{tabular}
\end{center}
\end{table}

\section{Methods}
\label{method}
We provide a concise overview of our methodology in the following subsections.
\subsection{Network embedding}
Network embedding is a technique in machine learning that transforms nodes in a network into low-dimensional vectors, capturing the structural and relational information of the graph. 
We employed the node2vec algorithm~\cite{grover2016node2vec} to embed each weekly weighted directed network into a $D$-dimensional space. Setting the return parameter $p$ and in-out parameter $q$ both to 1 in the algorithm signifies the use of unbiased random walks. Consequently, every node in the networks is represented by a $D$-dimensional vector, denoted as $V_i^\alpha$, where $i$ and $j$ are node indices, and $\alpha$ and $\beta$ are components in the $D$-dimensional space. This algorithm encodes both local community structure and global network topology in the learned node embeddings.

\subsection{Correlation tensor}
Utilizing the correlation tensor method and its diagonalization through double Singular Value Decomposition (SVD) offers valuable insights into crypto transaction systems.~\cite{chakraborty2023projecting, chakraborty2023dynamic, chakraborty2023embedding}.  
In the weekly XRP transaction networks, we consider $N$ nodes conducting at least one weekly transaction during the investigated period, categorizing them as regular nodes. Each regular node within the embedding space is presented as a time series of $D$-dimensional vectors, indicated as $V_{i}^\alpha (t)$. Here, $i$ varies from $1$ to $N$, $t$ spans from $1$ to $T$, and $\alpha$ ranges from $1$ to $D$.
The correlation tensor among components of regular nodes is expressed as:
\begin{equation}
M_{ij}^{\alpha\beta}(t) =\frac{1}{2\Delta T}\sum\limits_{t^\prime=t-\Delta T}^{t+\Delta T}\frac{[V_{i}^\alpha (t^\prime) - \overline{V_{i}^\alpha}][V_{j}^\beta (t^\prime) - \overline{ V_{j}^\beta}]}{\sigma_{V_i^\alpha} \sigma_{V_j^\beta}},
\label{eqn1}
\end{equation}
In this equation, we compute the sum across five weekly networks at times $t^\prime = \{t-2, t-1, t, t+1, t+2\}$ within a time window of $(2 \Delta T + 1)$, where $\Delta T=2$ for our analysis. The terms $\overline{V_{i}^\alpha}$ and $\sigma_{V_i^\alpha}$ denote the mean and standard deviation of $V_{i}^\alpha$ over a time window of $(2 \Delta T + 1) = 5$ weekly networks at times $\{t-2, t-1, t, t+1, t+2\}$. It's crucial to acknowledge that a smaller $\Delta T$ introduces more noise into the correlation tensor. However, opting for a larger $\Delta T$ is not feasible due to the detailed temporal evolution of the networks. In this analysis, we adopt a dimensionality of $D = 32$. The influence of the correlation tensor on window size $(2 \Delta T + 1)$ and dimension $D$ is elaborated in~\cite{chakraborty2023projecting}.

\subsection{Double singular value decomposition}

To get the spectrum of the correlation tensor, we employ a double SVD approach as outlined below:

We consecutively diagonalize $M_{ij}^{\alpha\beta}$ through a bi-unitary transformation, or SVD, first in terms of the $(ij)$-index and subsequently the $(\alpha\beta)$-index.
In the initial step, we represent $M_{ij}^{\alpha\beta}$ as a sum of matrices using the SVD technique.
\begin{equation}
M_{ij}^{\alpha\beta} = \sum\limits_{k=1}^N L_{ik}^{\alpha\beta}\sigma_k^{\alpha\beta} R_{kj}^{\alpha\beta}.
\label{eqn2}
\end{equation}
In the subsequent stage, we proceed to decompose each singular value $\sigma_k^{\alpha\beta}$ into a sum of matrices, employing the SVD method:
\begin{equation}
\sigma_k^{\alpha\beta} = \sum\limits_{\gamma=1}^D \mathcal{L}_k^{\alpha\gamma} \rho_k^\gamma \mathcal{R}_k^{\gamma\beta}.
\label{eqn3}
\end{equation}
Finally, substituting Eq.~\ref{eqn3} into  Eq.~\ref{eqn2},  we get the following expression for $M_{ij}^{\alpha\beta}$:
\begin{equation}
M_{ij}^{\alpha\beta} = \sum\limits_{k=1}^N \sum\limits_{\gamma=1}^D \rho_k^\gamma (L_{ik}^{\alpha\beta} R_{kj}^{\alpha\beta}) (\mathcal{L}_k^{\alpha\gamma} \mathcal{R}_k^{\gamma\beta}).
\label{eqn4}
\end{equation}
where $\rho_k^\gamma$ denotes the $N \times D$ generalized singular values, characterized by being both real and positive, due to all elements of the correlation tensor, $M$ is real.

\begin{figure}[tbh]
\includegraphics[width=0.98\textwidth]{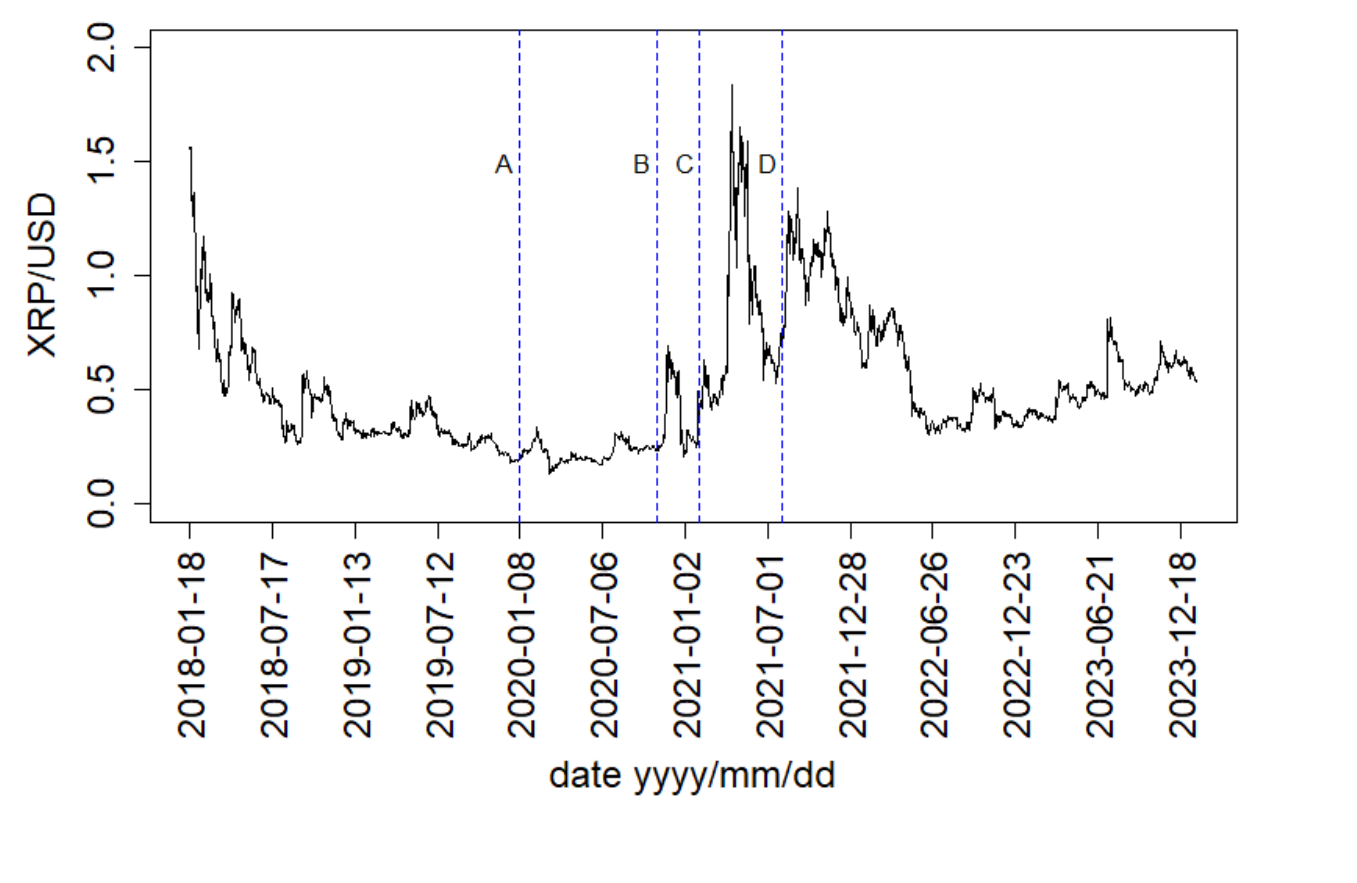}
\caption{The daily XRP prices at Huboi (HTX) crypto exchange. The four vertical lines represent the following dates: A) January 6, 2020; B) November 1, 2020; C) February 1, 2021; and D) August 1, 2021. }
\label{f1}
\end{figure}

\section{Results}
\label{result}

We show the daily XRP prices from January 18, 2018, to January 22, 2024, at the Huobi exchange in Fig.~\ref{f1}. The graph illustrates a considerable fluctuation in XRP prices, ranging from 0.136 USD to 1.838 USD during this timeframe. Vertical lines on the Fig. mark specific dates: A) January 6, 2020; B) November 1, 2020; C) February 1, 2021; and D) August 1, 2021. In this study, our primary focus lies on the periods AB and CD. These two phases exhibit distinct characteristics regarding XRP price fluctuations. Notably, the XRP price remained relatively stable during period AB.
In contrast, period CD showcases a notable surge and decline in XRP prices. We label CD as the "bubble period" and AB as the "non-bubble period" for XRP prices.

\begin{figure}[tbh]
\includegraphics[width=0.98\textwidth]{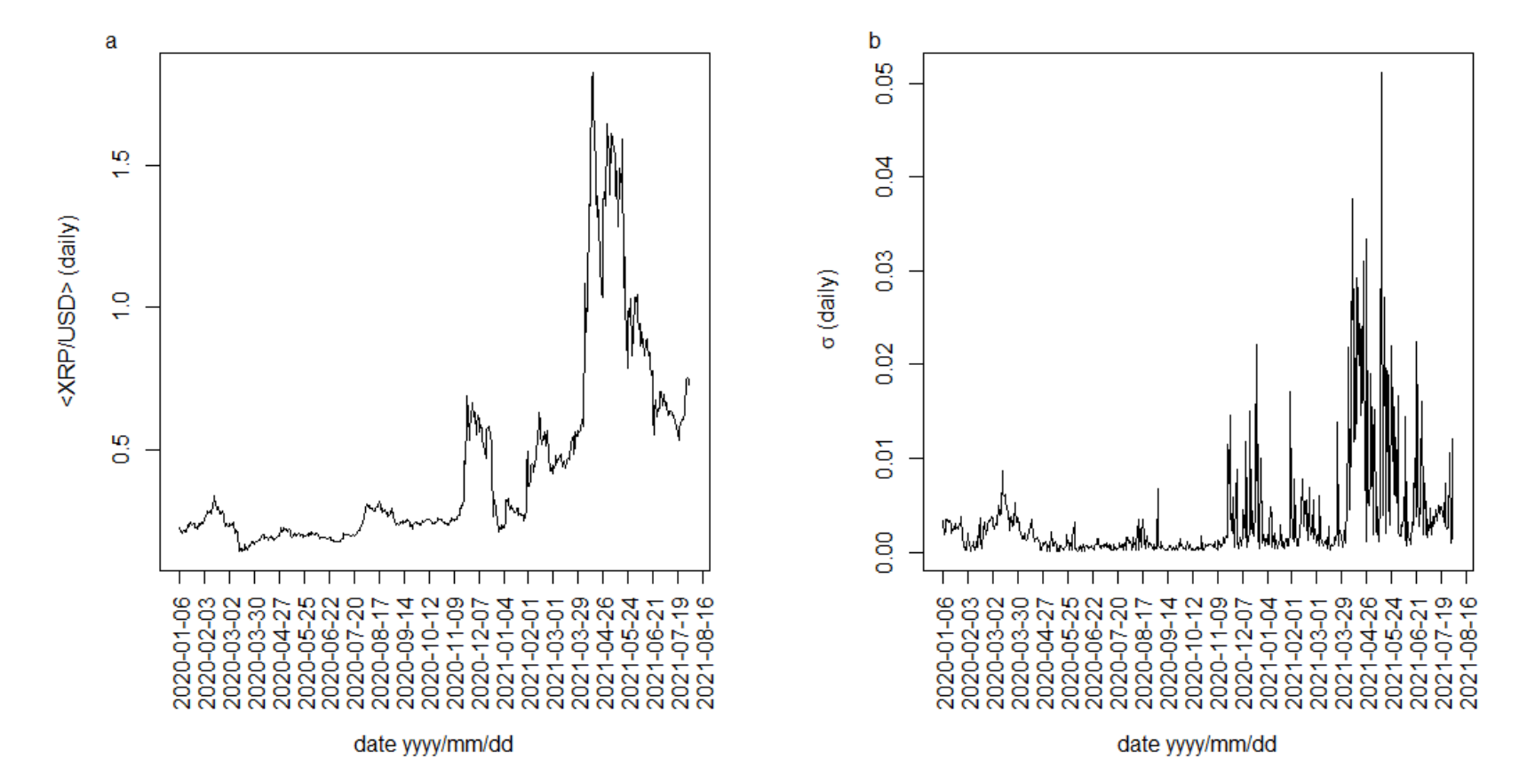}
\caption{Mean daily XRP prices, $\langle{\rm XRP/USD}\rangle$ (daily) and fluctuation $\sigma$ (daily)  over nine different crypto exchanges. }
\label{f2a}
\end{figure}

\begin{figure}[tbh]
\includegraphics[width=0.98\textwidth]{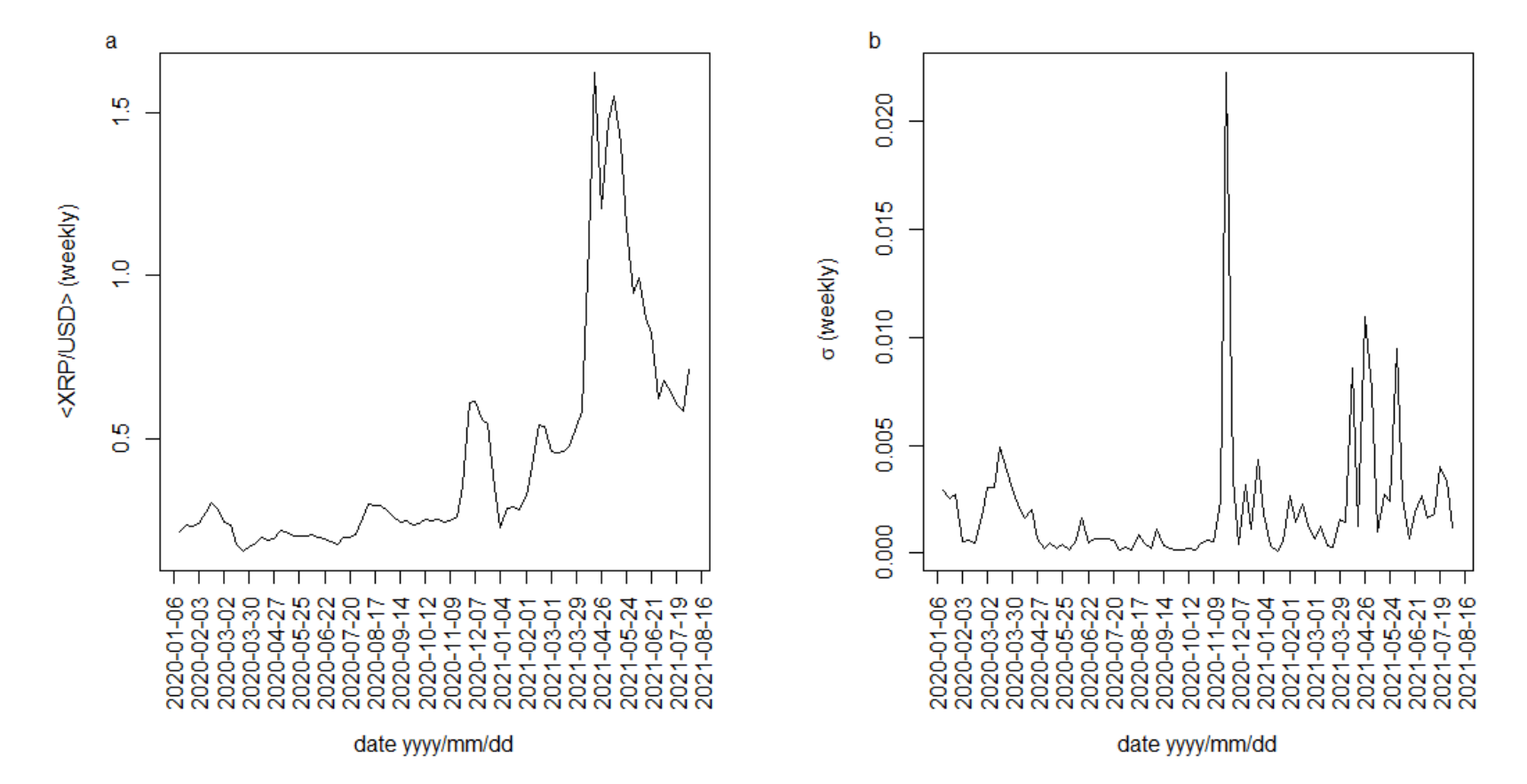}
\caption{Mean weekly XRP prices, $\langle{\rm XRP/USD}\rangle$ (weekly) and fluctuation $\sigma$ (weekly) over nine different crypto exchanges. }
\label{f2b}
\end{figure}

We examine how the XRP price varies among different crypto exchanges in Fig.~\ref{f2a}. We have considered daily XRP prices from nine crypto exchanges, which are listed in Table 1. Considering these nine crypto exchanges, we calculated the daily mean XRP price, $\langle{\rm XRP/USD}\rangle$ (daily), shown in Fig.~\ref{f2a}~(a) for the period AD. The daily fluctuation in XRP price among the exchanges is measured by the standard deviation $\sigma$. The fluctuation in the daily XRP price among the crypto exchanges is illustrated in Fig.~\ref{f2a}~(b).
The fluctuation in the daily XRP price among the crypto exchanges was very low during the first ten months of 2020. However, we observed a high fluctuation, even up to $5\%$, beyond the first ten months of 2020. We calculated the weekly XRP price, $\langle{\rm XRP/USD}\rangle$ (weekly), for a particular crypto exchange by averaging the daily XRP price over seven days. The mean $\langle{\rm XRP/USD}\rangle$ (weekly) and standard deviation $\sigma$ (weekly) in the weekly XRP price among the crypto exchanges are shown in Fig.~\ref{f2b}~(a) and (b), respectively. The fluctuations are relatively low for weekly XRP prices, as expected.

\begin{figure}[tbh]
\includegraphics[width=0.98\textwidth]{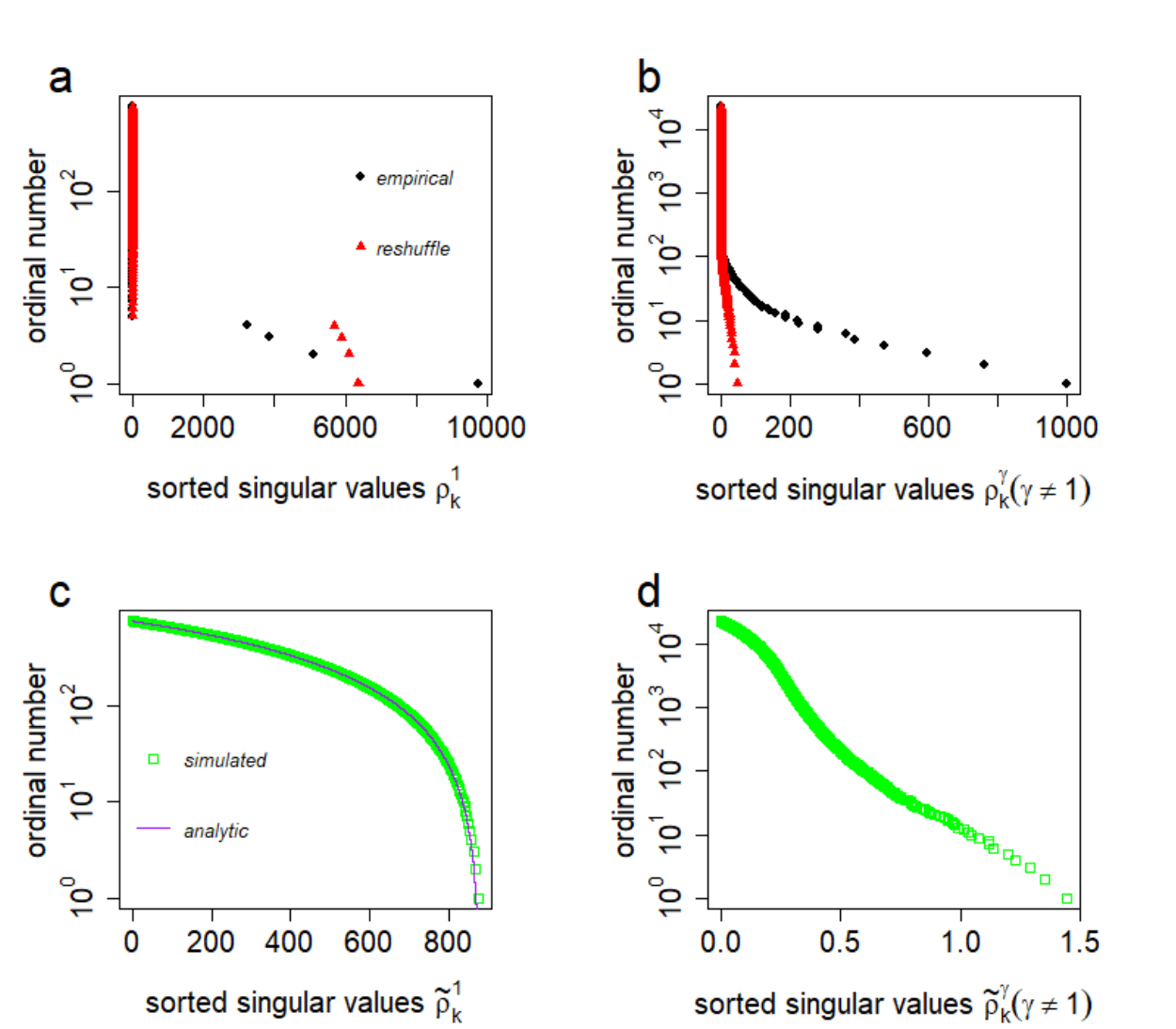}
\caption{ Singular value examination of correlation tensors from April 5 to 11, 2021.
(a) Comparative analysis of singular values, $\rho_k^1$, for the empirical correlation tensor (depicted by black filled circles) and the reshuffled correlation tensor (represented by red filled triangles) across all $k$ values.
(b) Exploration of singular values, $\rho_k^\gamma$, for the empirical correlation tensor and the reshuffled correlation tensor, encompassing all $k$ values with $\gamma > 1$.
(c) Simulated singular values, $\rho_k^1$, for the Gaussian random correlation tensor (illustrated as green open squares), accompanied by the corresponding analytic curve from Eq.~\ref{eqn8} (solid purple line).
(d) Examination of simulated singular values, $\rho_k^\gamma$, for the Gaussian random correlation tensor, considering all values of $k$ with $\gamma > 1$.
}
\label{f3}
\end{figure}

To establish a connection between the XRP price and the XRP transaction network, we compute the correlation tensor among the components of the regular nodes, denoted as $N$. This process is detailed in the Method section. The number of regular nodes is determined to be 465 and 753 for the time periods AB and CD, respectively. Focusing on the regular nodes during period CD, we compute the correlation tensor $M_{ij}^{\alpha\beta} (t)$ for the week $t =$ April 5-11, 2021.
The correlation tensor contains a total of $N \times N \times D \times D$ elements. Extracting essential insights from this tensor involves diagonalizing it through a technique known as double SVD, as detailed in Method section. The double SVD is an extension 
of the standard SVD, typically employed for matrices. By applying double SVD to the weekly correlation tensor, $M_{ij}^{\alpha\beta} (t)$, we derive the singular values denoted as $\rho_k^\gamma (t)$.

Determining the significance of the empirical correlation tensor requires comparing it with the reshuffled correlation tensor. To construct the reshuffled correlation tensor, we perform a reshuffling of the components within the time window $(2\Delta T +1)$ for the embedded regular node vector $V_i^\alpha$. Utilizing these reshuffled embedded regular node vectors, we compute the reshuffled correlation tensor according to Eq.~\ref{eqn1}. Furthermore, we simulate singular values of a Gaussian random correlation tensor utilizing random matrix theory~\cite{sengupta1999distributions, edelman2005random, bouchaud2009financial, rudelson2010non, bryc2020singular}. The elements of the Gaussian correlation tensor, denoted as $G_{ij}^{\alpha\beta}$, are drawn from a Gaussian distribution with a mean of zero and a standard deviation of $\sigma_G=0.5$, where $(i,j = 1, \ldots, N)$ and $(\alpha, \beta = 1, \ldots, D)$. We opt for $\sigma_G=0.5$ to align with the standard deviation of our empirical correlation tensor.

The probability distribution function for the largest singular values of the Gaussian random correlation tensor, denoted as $(\tilde{\rho}_k^1)$ for all $k$, is expressed as~\cite{chakraborty2023dynamic}:
\begin{equation}
P(\tilde{\rho}_k^1) = \frac{1}{\pi \sigma_G^2} \sqrt{(\tilde{\rho}_1^1)^2-(\tilde{\rho}_k^1)^2}. 
\label{eqn8}
\end{equation}
Here, the largest singular value is given by $\tilde{\rho}_1^1=2\sigma_G D \sqrt{N}$ for $k=1$.

\begin{figure}[tbh]
%\begin{center}
   \includegraphics[width=0.98\textwidth]{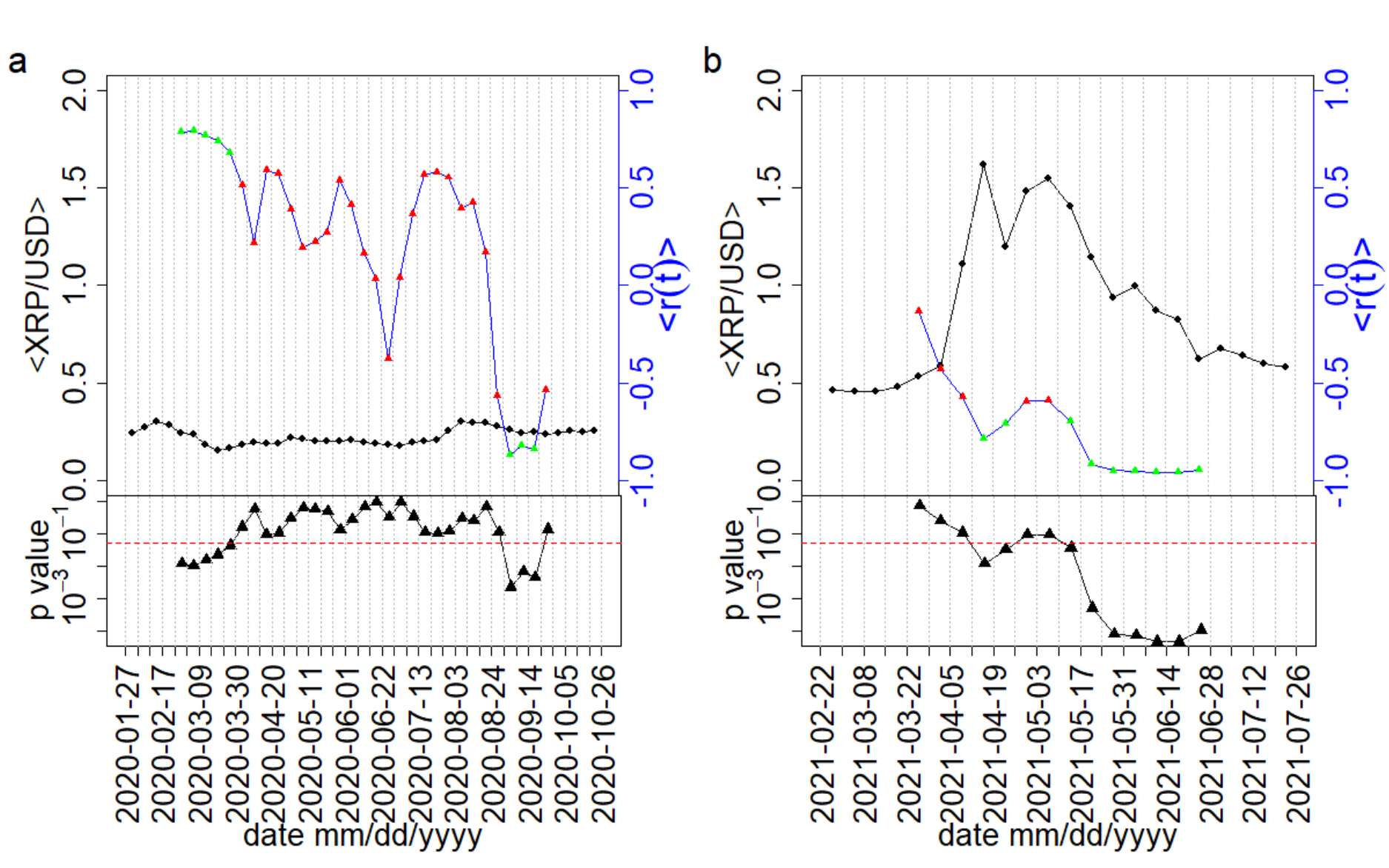}
\caption{Comparison of the mean simulated weekly XRP prices, $\langle {\rm XRP/USD (t)} \rangle$(simulated) with the mean correlation coefficient, $\langle r(t)\rangle$ , depicting the relationship between the mean simulated weekly XRP prices, $\langle {\rm XRP/USD (t)} \rangle$(simulated) and the largest singular value $\rho_1^1 (t-1)$. A moving window of 9 weeks is applied for two distinct periods: (a) AB, covering January 6, 2020, to November 1, 2020, and (b) CD, ranging from February 1, 2021, to August 1, 2021.
The black curves illustrate the mean simulated weekly XRP prices, $\langle {\rm XRP/USD (t)} \rangle$(simulated). The blue curves, marked with green and red triangles, represent the mean correlation coefficient, $\langle r(t)\rangle$, with green triangles denoting significant correlations (p-value $< 0.05$) and red triangles indicating no significant correlations (p-value $> 0.05$). The two lower panels display the corresponding p-values for Pearson correlations. Dotted grey vertical lines delineate the weekly windows. 
}
%\end{center}
\label{ff4}
\end{figure}

Fig.\ref{f3}~(a) displays the singular values $\rho_k^\gamma$ for all $k \in [1,N]$ and $\gamma=1$, along with the singular values for the reshuffled correlation tensor. Notably, only the largest singular value $\rho_1^1$ for the empirical correlation tensor surpasses the largest singular value of the reshuffled correlation tensor.

We extend this comparison to other singular values $\rho_k^\gamma$ for all $k \in [1,2,3, \dots, N]$ and $\gamma \in [2,3,4, \dots, D]$ in Fig.\ref{f3}~(b). Here, we observe that several singular values of the empirical correlation tensor surpass the largest singular values of the reshuffled correlation tensor, although these singular values are notably smaller than $\rho_1^1$. Consequently, their contribution to the correlation tensor is relatively minor.

Additionally, we compare the empirical singular values with those of the Gaussian random correlation tensor, where the elements are drawn from a normal distribution. The singular values $\tilde{\rho}_k^1$ of the Gaussian random correlation tensor follows Eq.~\ref{eqn8}.
In Fig.\ref{f3}(c), the simulated singular values, $\tilde{\rho}_k^1$, of the Gaussian correlation tensor exhibit a smooth fit with the analytical curve provided by Eq.\ref{eqn8}.
 Fig.\ref{f3}~(d) presents the spectrum, $\tilde{\rho}_k^\gamma$, of the Gaussian random correlation tensor across all $k$ values, where $\gamma$ ranges from 2 to $D$. It is evident that the singular values of the Gaussian correlation tensors are much smaller than those of the empirical and reshuffled correlation tensors. The singular values of the reshuffled correlation tensor tend to approach those of the Gaussian correlation tensor when the time window $\Delta T$ becomes significantly larger than $N$ ~\cite{chakraborty2023projecting}. While we present results specifically for the week April 5-11, 2021, it is worth noting that these findings remain qualitatively consistent for any other week.

We generate a set of $1000$ time series representing weekly XRP prices, denoted as ${\rm XRP/USD}(t)$ (simulated). These time series are simulated from Gaussian distributions characterized by the mean $\langle{\rm XRP/USD}\rangle$ (weekly) and the standard deviation $\sigma (t)$ (weekly). To assess the relationship between the simulated weekly XRP prices and the largest singular value $\rho_1^1 (t-1)$, we employ the Pearson correlation coefficient. Specifically, we calculate the correlation within a moving time window of length $9$ weeks.
The Pearson's correlation coefficient, denoted as $r(t)$, is defined as follows: 
\begin{equation}
r(t) =\frac{1}{2\Delta \tau}\sum\limits_{t^\prime=t-\Delta \tau}^{t+\Delta \tau}\frac{[y(t^\prime) - \langle y\rangle][\rho_1^1(t^\prime -1) - \langle \rho_1^1 \rangle]}{\sigma_{y} \sigma_{\rho_1^1}},
\label{eqn9}
\end{equation}
where the variable $y$ corresponds to ${\rm XRP/USD}(t)$ (simulated), and we have chosen $\Delta \tau = 4$. Herein, the symbols $\sigma$ and $\langle \cdot \rangle$ denote the standard deviation and mean, respectively, of the quantities within the temporal span of $(2 \Delta \tau + 1)$ weeks.

\begin{figure}[tbh]
\includegraphics[width=0.98\textwidth]{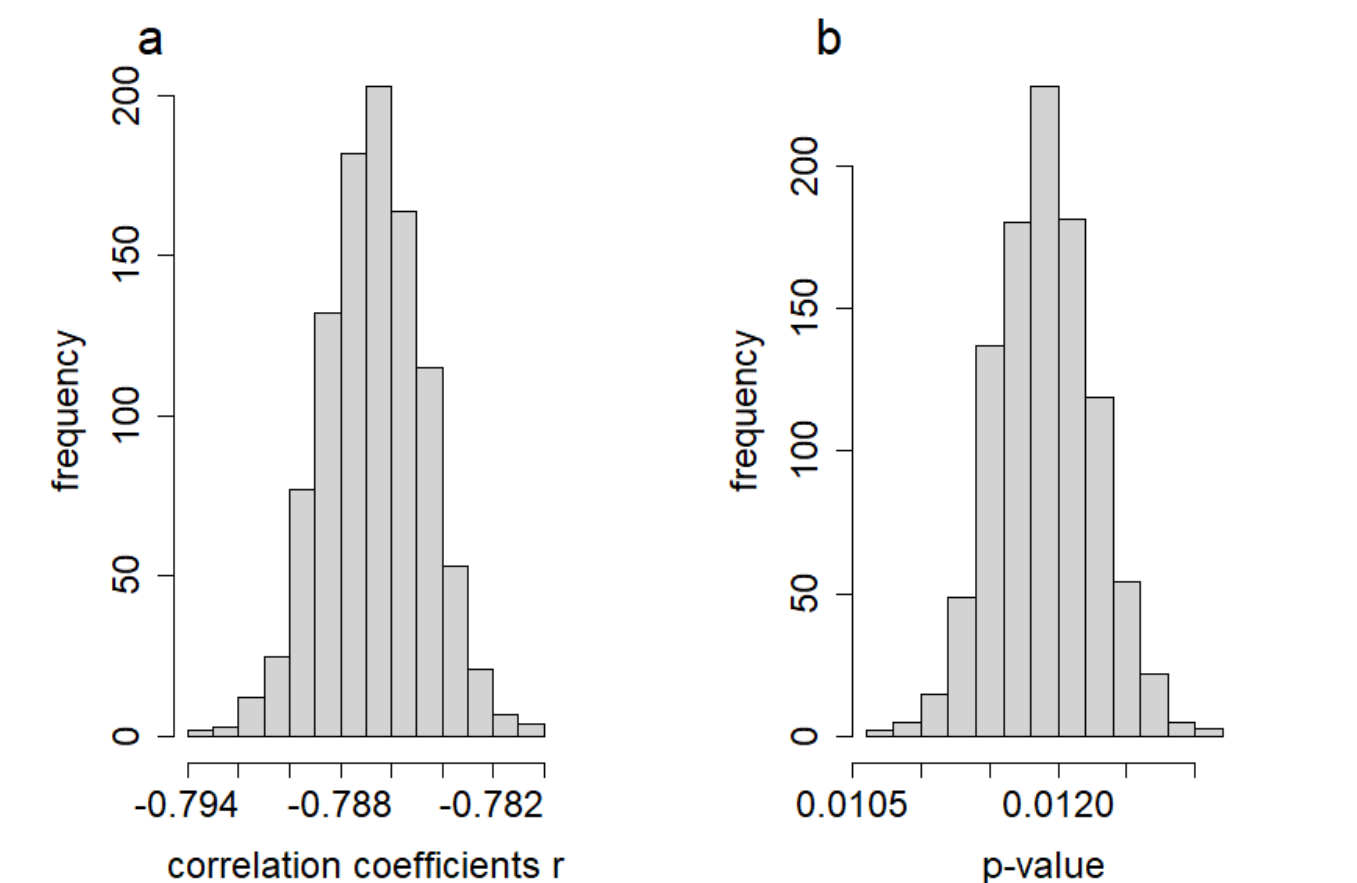}
\caption{(a) Distribution of correlation coefficients between the largest singular value $\rho_1^1(t-1)$ and simulated $1000$ weekly XRP Price times series $XRP/USD(t)$ for the week, $t= $April $12 -18, 2021$.
(b)Distribution of associated p-values. }

\label{f5}
\end{figure}

\begin{figure}[tbh]
\includegraphics[width=0.98\textwidth]{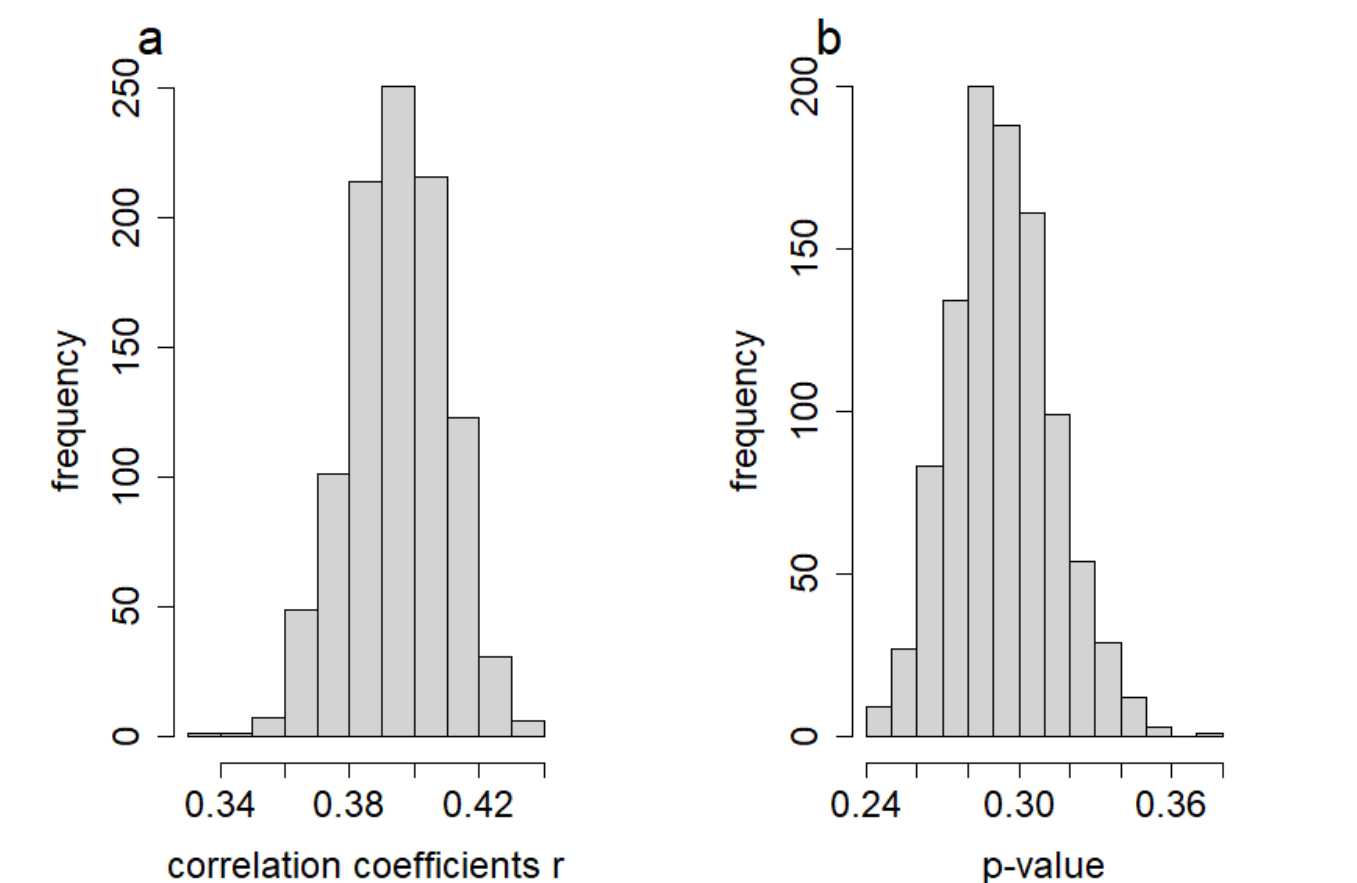}
\caption{(a) Distribution of correlation coefficients between the largest singular value $\rho_1^1(t-1)$ and simulated $1000$ weekly XRP Price times series $XRP/USD(t)$ for the week, $t= $April $27$ - May $3, 2020$.
(b)Distribution of associated p-values. }

\label{f6}
\end{figure}
We calculate mean  correlation coefficient $\langle r(t) \rangle$ over $1000$ simulated set of  weekly XRP prices, ${\rm XRP/USD}(t)$ (simulated). The mean simulated weekly XRP prices are represented as $\langle{\rm XRP/USD}(t)\rangle$ (simulated). We  investigate the temporal correlation separately for two different periods, AB and CD.
The temporal variation of the mean correlation coefficient $\langle r(t) \rangle$ between the simulated weekly XRP prices, ${\rm XRP/USD}(t)$ (simulated), and the largest singular values $\rho_1^1 (t-1)$ of the weekly correlation tensors is shown along with the  mean simulated weekly XRP prices, $\langle{\rm XRP/USD}(t)\rangle$ (simulated), in Fig.~\ref{ff4}. The anti correlation is high and significant during the period CD. The anti correlation mostly non-significant during the period AB. This reflects the fact that the formation of a large bubble in the XRP price is indicated by a strong anti correlation with the largest singular value.  The findings underscore the impact of XRP price fluctuations across various exchanges. Specifically, the largest singular value, $\rho_1^1$, demonstrates a pronounced and meaningful average negative correlation during the bubble period. Conversely, during non-bubble periods, the average correlation is observed to be non-significant.

Our analysis extends to a more detailed exploration of the correlation coefficients with each simulated weekly XRP price time series, ${\rm XRP/USD}(t)$ (simulated). Specifically, we focus on two distinct weeks, namely $t_1$ and $t_2$, each chosen from distinct periods - $t_1$ from April 12-18, 2021, falling within the bubble period CD, and $t_2$ from April 27- May 3, 2020, situated within the non-bubble period $AB$.
In Fig.~\ref{f5}, we present the distributions of correlation coefficients, $r(t)$, along with associated p-values between the largest singular value $\rho_1^1(t-1)$ and a simulated set of 1000 weekly XRP Price time series, ${\rm XRP/USD}(t)$ (simulated), for the week $t= t_1$. Notably, all correlation coefficients $r(t)$ exhibit a highly negative and statistically significant pattern.
Conversely, in Fig.~\ref{f6}, we illustrate the distributions of correlation coefficients $r(t)$ and corresponding p-values for the same analysis conducted for the week $t= t_2$. Remarkably, all correlation coefficients $r(t)$ in this case are observed to be non-significant. These findings provide additional affirmation that our results remain consistent across each simulated time series. This underscores the notion that price fluctuations across various crypto exchanges do not exert any significant influence on our conclusions. This phenomenon can be attributed to the minimal price variations of XRP across various trading exchanges, leading to limited arbitrage opportunities. Notably, the p-values associated with the correlation between the largest singular value of the correlation tensor and the price reveal that this correlation remains significant even when accounting for the price differentials of XRP across diverse exchanges, particularly during the bubble period.

\section{Conclusions}
\label{conclusion}

We have examined the impact of XRP price fluctuations across different crypto exchanges on the correlation between the largest singular values of the correlation tensor of XRP transaction networks and the XRP price. For this purpose, we have collected XRP price data from nine different crypto exchanges. We have displayed the daily and weekly prices and their fluctuations over a period of approximately 1.5 years. To uncover the correlation between price and XRP transaction networks, we have calculated the singular values of correlation tensor spectra. Comparing this with a reshuffled correlation tensor, we found that only the largest singular value is significant.

To capture the price impact of fluctuations, we have generated a set of 1000 simulated weekly XRP price time series from a Gaussian distribution with an identical mean and standard deviation as empirical XRP prices. We have found that irrespective of XRP price fluctuations at crypto exchanges, the largest singular value shows a strong, significant anti-correlation during the bubble period and a non-significant correlation during the non-bubble period with the XRP price. Our simulated weekly XRP price mirrors the potential arbitrage opportunities present across various cryptocurrency exchanges. Upon examination of correlation coefficients and p-values between the largest singular value and these time series, it becomes evident that arbitrage exerts a non-significant impact. This can be attributed to the relatively low arbitrage opportunities across diverse cryptocurrency exchanges.

{
\appendix

\section{Cryptoassets monthly exchange volume}
We illustrate the monthly spot market volume across three prominent cryptoasset exchanges from January 2020 to August 2020 in Fig.~\ref{f7}. It visually demonstrates the evolution of Binance as it ascends to become the largest crypto exchange during this period.
\begin{figure}[tbh]
\includegraphics[width=0.98\textwidth]{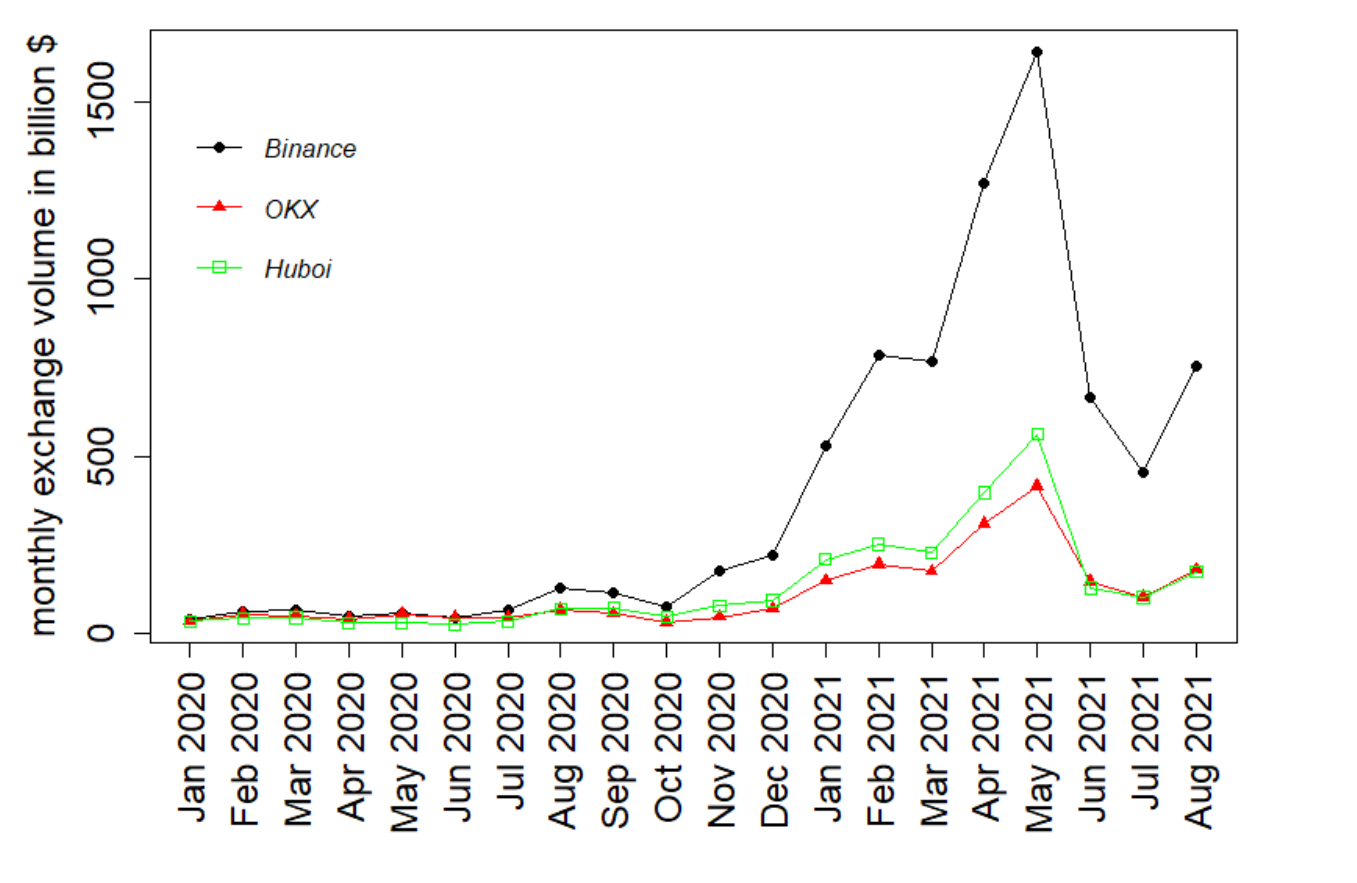}
\caption{Monthly spot market volumes across three major cryptoasset exchanges from \url{https://www.theblock.co/data/crypto-markets/spot}.  }
\label{f7}
\end{figure}

\section{Data sources:}
\begin{itemize}
%\item Investing.com: \url{https://www.investing.com/crypto/xrp/historical-data}
\item Binance: \url{https://www.investing.com/crypto/xrp/xrp-btc-historical-data}
\item OKX: \url{https://www.investing.com/crypto/xrp/xrp-usd-historical-data}
\item Huboi: \url{https://www.investing.com/crypto/xrp/xrp-usd-historical-data?cid=1058260}
%\item Bitrue: \url{https://www.investing.com/crypto/xrp/xrp-usd-historical-data?cid=1179839}
\item Kraken: \url{https://www.marketwatch.com/investing/cryptocurrency/xrpusd}
\item Bitstamp: \url{https://www.CryptoDataDownload.com}
\item Bitfinex: \url{https://www.CryptoDataDownload.com}
\item Exmo: \url{https://www.CryptoDataDownload.com}
\item Bittrex: \url{https://www.CryptoDataDownload.com}
\item FTX: \url{https://www.CryptoDataDownload.com}
\item Transaction networks: \url {https://xrpl.org/data-api.html#payment-objects}.
\end{itemize}
}
\section*{Acknowledgements}
We express our gratitude to Tetsuo Hatsuda for his valuable comments on our manuscript. 
%We also thank the members of the Kyoto Univ.- RIKEN blockchain study group  for discussions.   
YI acknowledges the financial support received from the Ripple Impact Fund (Grant Number: 2022-247584) for partial support to this work. 

%\begin{figure}[tbh]
%\includegraphics[width=0.98\textwidth]{figures/SV_price.png}
%\caption{Analyzing the daily XRP/USD price alongside singular values and spectral gap from June 6, 2020, to July 4, 2021, denoted as period $XX$.
%The graph features black curves representing the daily XRP/USD price, while the blue curves depict (a) the largest singular value $\rho_1^1$,
%(b) the second largest singular value $\rho_2^1$, and (c) the spectral gap $(\rho_1^1-\rho_2^1)$ of correlation tensors across different weeks. 
%}
%\label{f3}
%\end{figure}

%\clearpage
%\newpage

%
%\section{}

%Use the \verb|\appendix| command if you need an appendix(es). The \verb|\section| command should follow even though there is no title for the appendix (see above in the source of this file).
%\bibliographystyle{apsrev4-1}
%\bibliography{bibl}
%\newpage

%

\end{document}